\documentclass{aa}
\input epsf.tex
\begin{document}

\thesaurus{08.19.4;09.02.1;10.05.1;10.07.2}

\title{The self-enrichment of galactic halo globular clusters :  }
 
\subtitle{a clue to their formation ?}

\author{G. Parmentier, 
        E. Jehin, 
        P. Magain\thanks{Ma\^{\i}tre de Recherches au 
                 Fonds National de la Recherche Scientifique (Belgium)}, 
        C. Neuforge, 
        A. Noels and 
        A.A. Thoul\thanks{Chercheur qualifi\'e au 
                    Fonds National de la Recherche Scientifique (Belgium)}}   
\offprints{G.\ Parmentier}

\institute{Institut d'Astrophysique et de G\'eophysique, Universit\'e de 
           Li\`ege, 5, Avenue de Cointe, B-4000 Li\`ege, Belgium}
\date{Received 14 June 1999/ Accepted 18 October 1999}
\maketitle
\markboth{G.\ Parmentier et al.: 
          The self-enrichment of galactic halo globular clusters}{}

\begin{abstract}

We present a model of globular cluster self-enrichment.  In the protogalaxy, 
cold and dense clouds embedded in the hot 
protogalactic medium are assumed to be the progenitors of galactic halo 
globular clusters.
The massive stars of a first generation of metal-free stars, born in 
the central areas of the proto-globular cluster clouds, explode as Type 
II supernovae.  The associated blast waves trigger the expansion of 
a supershell, sweeping 
all the material of the cloud, and the heavy elements released by these 
massive stars enrich the supershell.  A second generation of stars is 
born in these compressed and enriched layers of gas.  These stars 
can recollapse and form a globular cluster.  
This work aims at revising the most often encountered argument against
self-enrichment, namely the presumed ability of a small number of supernovae to
disrupt a proto-globular cluster cloud. 
We describe a model of the dynamics of the supershell and of its 
progressive chemical enrichment.  
We show that the minimal mass of the primordial cluster cloud 
required to avoid disruption by several tens of Type II supernovae is 
compatible with the masses usually assumed for
proto-globular cluster clouds.  Furthermore, 
the corresponding self-enrichment level is
in agreement with halo globular cluster metallicities.

\keywords{globular clusters: general -- Galaxy: evolution -- 
supernovae: general -- ISM: bubbles -- population III}

\end{abstract}

\section{Introduction}

The study of the chemical composition and dynamics of the galactic halo 
components, field metal-poor stars and globular clusters (hereafter GCs), 
provides a 
natural way to trace the early phases of the galactic evolution.  In an 
attempt to get some new insights on the early galactic nucleosynthesis, 
accurate relative abundances have been obtained from the analysis of
high resolution  and high signal-to-noise spectra for a sample of 21
mildly metal-poor stars (Jehin et al.~1998, 1999).
The correlations between the relative
abundances of 16 elements have been studied, with a special emphasis on the
neutron capture ones.  This analysis reveals the existence of two
sub-populations of field metal-poor stars, namely Pop IIa and Pop IIb.  They
differ by the behaviour of the $s$-process elements versus the $\alpha$ and
$r$-process elements.
To explain such correlations, Jehin et al.~(1998, 1999) have suggested
 a scenario for the formation of metal-poor stars which closely 
relates  the origin of these stars to the formation and the evolution of 
galactic globular clusters. \\
At present, there is no widely accepted theory of globular cluster formation.
According to some scenarios,
GC formation represents the high-mass tail of star cluster 
formation.  Bound stellar clusters form in the dense cores of much 
larger star-forming clouds with an efficiency of the order $10^{-3}$ to 
$10^{-2}$ (Larson, 1993).  
If GCs form in a similar way, the total 
mass of the protoglobular clouds should therefore be two or three orders of 
magnitude greater than the current GC masses, leading to a total mass of 
$10^8$M$_{\odot}$.  Harris and Pudritz (1994) have investigated the 
GC formation in such clouds which they call SGMC 
(Super Giant Molecular Clouds).  The physical conditions in these SGMC 
have been further explored by McLaughlin and Pudritz (1996a).  \\    
Another type of scenarios rely on a 
heating-cooling balance to preserve a given temperature 
(of the order of $10^4$K) 
and thus a characteristic Jeans Mass at the protogalactic epoch.
In this context, Fall and Rees (1985) propose that GCs would form in the 
collapsing gas of the protogalaxy.  During this collapse, a thermal 
instability triggers the development of a two-phase structure, namely cold 
clouds in pressure equilibrium with a hot and diffused medium.  
They assume that the temperature of the cold clouds remains at
$10 ^4$K since the cooling rate drops sharply at this temperature in a 
primordial gas.  This 
assumption leads to a characteristic mass of order $10^6$
M$_{\odot}$ for the cold clouds.  
However, this temperature, and therefore the characterictic mass, 
is preserved only if there is a flux of UV or X-ray radiation able 
to prevent any $\rm H_2$ formation, the main coolant in a metal-free gas below 
$10^4$K. 
Provided that this condition is fullfilled, and since the characteristic 
mass is of the order of GC masses (although a bit larger, but see 
section 5.2.1), Fall and Rees identify the cold clouds with the progenitors 
of GCs.  Several formation scenarios have included their key idea : 
cloud-cloud collisions (Murray and Lin 1992), self-enrichment model 
(Brown, Burkert and Truran 1991, 1995).  \\
According to the scenario suggested by Jehin {\it et al.} (1998, 1999), 
GCs may have undergone a Type II supernovae phase in their early history.
This scenario appears therefore to be linked with the self-enrichment 
model developed by Brown {\it et al.} within the context of the Fall 
and Rees theory.
Following Jehin {\it et al.}, thick disk and field halo stars were born in 
globular clusters from which they escaped either during an early disruption 
of the proto-globular cluster (Pop IIa) or through a later disruption 
or evaporation process of the cluster (Pop IIb). The basic idea is that
the chemical evolution of the GCs can be described in two phases.
During phase I, a first generation of metal-free stars form in the 
central regions of proto-globular cluster clouds (hereafter PGCC).  The 
corresponding massive stars 
evolve, end their lives as Type II supernovae (hereafter SNeII) and 
eject $\alpha$, $r$-process and possibly a small amount of light 
$s$-process elements into the interstellar medium.  A second generation 
of stars form out of this enriched ISM.  If the PGCC get disrupted, 
those stars form Pop IIa.  If it
survives and forms a globular cluster, we get to the second phase where 
intermediate mass stars reach the AGB stage of stellar evolution, ejecting 
$s$ elements into the ISM through stellar 
winds or superwind events.  The matter released in the ISM by AGB stars 
will be accreted by lower mass stars, enriching their external layers  
in $s$ elements.  During the subsequent dynamical evolution of the globular 
cluster, some of the surface-enriched low-mass stars evaporate from the 
cluster, become field halo stars and form PopIIb. \\
Others studies have already underlined the two star generations concept : 
Cayrel (1986) and Brown et al.~(1991, 1995) were pioneers in this field.  
Zhang and Ma (1993) have demonstrated that no single star formation can fit 
the observations of GCs chemical properties.  They show that there must be 
two distinct stages of star formation : a self-enrichment stage (where 
the currently 
observed metallicity is produced by a first generation of stars) and a 
starburst stage (formation of the second generation stars). \\
These two generations scenarios were marginal for a long time. 
Indeed, the major criticism of such globular cluster self-enrichment model 
is based
on the comparison between the energy released by a few supernova explosions 
and the PGCC gravitational binding energy : they are of the same order of 
magnitude.  It might seem, therefore, that proto-globular cluster clouds 
(within the context of the Fall and Rees theory)
cannot survive a supernova explosion phase and are disrupted (Meylan and 
Heggie, 1997). Nevertheless, 
a significant part of the energy released by a supernova 
explosion is lost by radiative cooling (Falle, 1981) and the kinetic 
energy fraction interacting with the ISM must be reconsidered. \\
While Brown et al.~(1991, 1995) have mostly focused on the computations of
the supershell behaviour through hydrodynamical computations, we 
revisit some of the ideas that have been used against the hypothesis 
of GC self-enrichment.  In this first paper, we tackle 
the questions of the supernova energetics and of the narowness of GC 
red giant branch. 
Dopita and Smith (1986) have already addressed the first point from a purely 
dynamical point of view.  In their model,
they assume the simultaneity of central supernova explosions and they use
the Kompaneets (1960) approximation to describe the resulting blast
wave motion.  During this progression from the central regions to the edge of 
the PGCC, all the material encountered by the blast wave is swept up into a 
dense shell.  
They demonstrate that, when the shell emerges from the cloud, its kinetic 
energy, based on the number 
of supernovae that have exploded, is compatible with the gravitational 
binding energy of a cloud whose mass is more or less $10^7M_{\odot}$.
When the kinetic energy of the emerging shell is larger than the 
binding energy of the initial cloud, this cloud is assumed to be disrupted by 
the SNeII.  There is therefore a relation 
between the cloud mass and the maximum number of supernovae 
it can sustain without being disrupted.  However, a $10^7M_{\odot}$ cloud
is more massive than the PGCCs considered within the Fall and Rees theory. \\
We derive here a similar relation based on 
the supershell description (Castor, McCray, Weaver, 1975)
of the central supernova explosions.
Contrary to the Kompaneets approximation, this theory allows us to take into
account the existence of a mass spectrum for the massive supernova 
progenitors and, therefore, the spacing in time of the explosions. 
In addition to the above dynamical constraint, we also establish a chemical 
one.
For a given mass of primordial gas, we compute the maximum number of supernovae
the PGCC can sustain and the corresponding self-enrichment level
at the end of the supernova phase.  We show that the metallicity 
reached is compatible with the metallicity observed in 
galactic halo globular clusters. \\
The paper is organized as follows.  In section 2, we review the observations 
gathered by Jehin et al.~(1998, 1999) and the scenario proposed 
to explain them.
In section 3, we describe the PGCCs, the first 
generation of metal-free stars, and the supershell propagation
inside PGCCs due to SNeII explosions.  In section 4, we 
show that the disruption criterion proposed by Dopita and Smith (1986), here 
computed with the supershell theory, gives the correct 
globular cluster metallicities.  In section 5, we discuss
the sensitivity of our model to the first generation 
IMF parameter values and we examine the implications of an important 
observational constraint, the RGB narrowness noticed in most globular cluster 
Color Magnitude Diagrams (CMDs).
Finally, we present our conclusions in section 6.

\section{The EASE scenario}

\subsection{Observational results}

Jehin et al.~(1998, 1999) selected a sample of 21 unevolved metal-poor stars 
with roughly one 
tenth of the solar metallicity.  This corresponds to the transition between 
the halo and the disk.  All stars are dwarfs or subgiants,
at roughly solar effective temperature and covering a narrow metallicity range.
High quality data have been obtained and a careful spectroscopic
analysis was carried out.  The scatter in element abundances 
reflects genuine cosmic scatter and not observational uncertainties.   
Abundances of iron-peak elements (V, Cr, Fe, Ni),
$\alpha$ elements (Mg, Ca, Ti), light $s$-process elements (Sr, Y, Zr), 
heavy $s$-process elements (Ba, La, Ce), an $r$-process element (Eu) and 
mixed $r$-,$s$-process elements (Nd, Sm) have been determined.  
Among these data, Jehin et al.~(1998, 1999) have found  
correlations between abundance ratios at a given metallicity.  
If some elements are 
correlated, they are likely to have been processed in the same astrophysical 
sites, giving fruitful information about nucleosynthesis.
The following results were obtained (Jehin et al.~1999) : 

\begin{itemize}

\item[a. ] there is a one-to-one correlation between the $r$-process 
element Eu and
the $\alpha$ element Ti : most points are located on a single straight  line
with a slope $ \simeq 1 $  ending with a clumping at the maximum value of
[Ti/Fe]

\item[b. ] looking for correlations including $s$-process element 
(e.g.~yttrium), two distinct behaviours were found : some 
stars show a correlation between [Y/Fe] and [Ti/Fe] with a slope smaller
than one while some others stars have a  constant (and maximum)
[Ti/Fe] value and varying values of [Y/Fe], starting at the  maximum 
value reached by the first group.
This vertical branch is the counterpart of the previous clump. 
We have labeled this diagram the \it two-branches diagram\rm (Fig.~1).
Since our sample contains a limited number of stars, we have added results
from other analysis concerning metal-poor stars ($-2 <$ [Fe/H] $ < -0.6$).
All these metal-poor stars follow the same trend independently of their 
metallicity. But if we now add results for disk stars of various 
metallicities, the relation obtained for metal-poor stars is no more verified 
: the points scatter in the upper-left part of the diagram. 
There is a change of behaviour at a metallicity [Fe/H]$ \simeq -0.6 $.  
The \it two-branches diagram \rm describes a relation that applies only to 
metal-poor stars.  Higher metallicity stars do not fit the \it two-branches 
diagram \rm .  
\end{itemize}

\begin{figure}
\begin{center}
\leavevmode 
\epsfxsize= 8.5 cm 
\epsffile{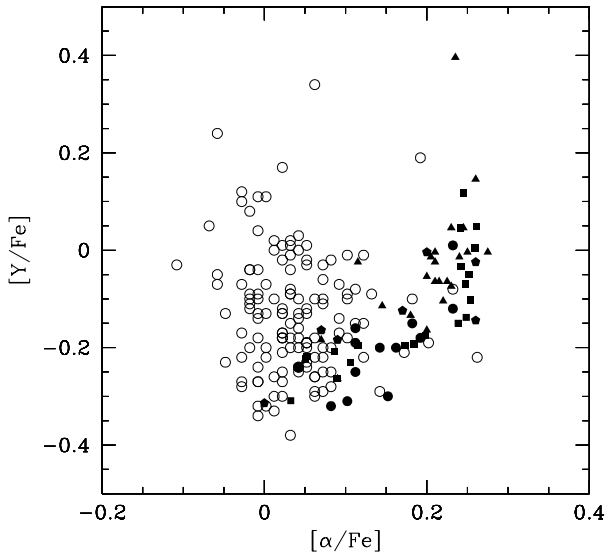}
\vspace*{-0.4 cm}
\caption{Correlation diagram for [Y/Fe] versus [$\alpha$/Fe] with the 
data from Jehin et al.~(1999) (full squares), the data of Nissen and 
Schuster (1997) (full triangles), the data of Zhao and Magain (1991) 
(full pentagon), the data of Edvardsson et al.~(1993) for [Fe/H]$<-$0.6 
(full circles) and for [Fe/H]$>-$0.6 (open circles).
The zero points have been fixed by comparing the results obtained for
the stars in common with Jehin et al.~(1999)}
\end{center}
\end{figure}

\subsection{Interpretation : two-phases globular cluster evolution}

We associate the two branches of the diagram with two distinct chemical 
evolution phases of globular clusters, namely a SNII phase 
(phase I) and an AGB wind phase (phase II). \\

\vspace{1ex}
\noindent
{\em Phase 1 } 

We assume the formation of a first generation of stars in the central regions,
the densest ones, of a 
proto-globular primordial gas cloud.  The most massive stars of this first
generation evolve and become supernovae, ejecting $\alpha$ and $r$-process 
elements into the surrounding ISM.  These supernova  explosions also trigger
the formation of an expanding shell, sweeping all the PGCC material encountered
during its expansion and decelerated by the surrounding ISM.
In this supershell, the supernova ejecta mix with the ambient ISM, 
enriching it in $\alpha$ and $r$-process elements.
The shell constitutes a dense medium since it contains all the PGCC gas in a 
very thin layer (a few tenths of parsecs) (Weaver et al., 1977).
This favours the birth of a second generation of 
stars (\it triggered \rm star formation) with a higher star formation 
efficiency (hereafter SFE) than the first 
one (\it spontaneous \rm star formation). 

\begin{figure}
\vspace*{0.3cm}
\begin{center}
\leavevmode 
\epsfxsize= 7.8 cm 
\epsffile{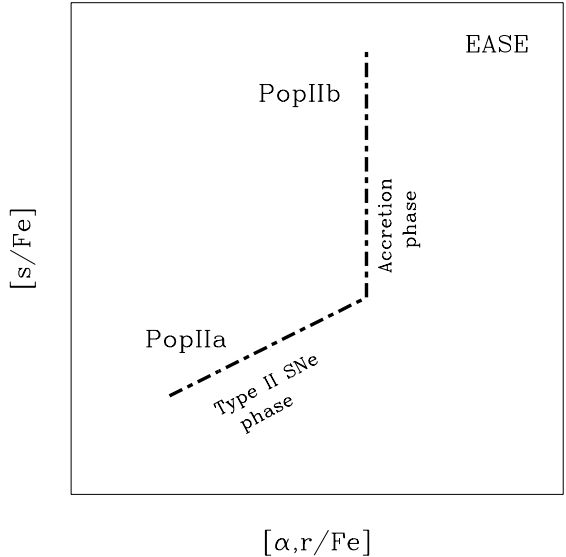}
\caption{EASE scenario (Jehin et al., 1999)}
\end{center}
\end{figure}

There are today several observational evidence of triggered star formation
(Fich, 1986; Walter et al., 1998; Blum et al., 1999). 
For instance, within the violent interstellar medium of the nearby dwarf 
galaxy IC2574,
Walter et al.~(1998) have studied a supergiant HI shell obviously produced 
by the combined effects of stellar winds and supernova explosions.
A major star formation event (equivalent to our first generation) has 
likely taken
place at the center of this supershell and the most massive stars have 
released energy into the ISM.  This one, swept by the blast waves associated 
with the first supernova explosions, accumulates in the form of an expanding 
shell surrounding  a prominent HI hole.
On the rim of the HI shell, H$\alpha$ emissions reveal the existence of 
star forming regions (equivalent to our second generation).  
Actually, these regions of HI accumulated by the sweeping 
of the shell have reached densities high enough for a secondary star 
formation to start via gravitational fragmentation. \\ 
In our scenario, the second generation of stars formed in this dense 
and enriched shell makes up the proto-globular cluster.  If the shell doesn't 
recontract, these stars 
form PopIIa and they appear somewhere on the slowly varying branch 
depending on the time at which the second generation formation has occurred.
When all stars more massive than $9 M_{\odot}$ have exploded as supernovae,
the $\alpha$ and $r$ element synthesis stops, leading to a typical value of 
[$\alpha$/Fe].
Our scenario requires this typical value of [$\alpha$/Fe] to be the maximum
value observed in the \it two-branches diagram \rm : the end of the supernova
 phase must correspond to the bottom of the vertical branch.
If the proto-globular cluster survives the supernova phase, 
the shell of stars will recontract and form a globular cluster. \\

\vspace{1ex}
\noindent
{\em Phase 2}

After the birth of the second generation, intermediate mass stars evolve until
they reach the asymptotic giant branch where they enrich their envelope 
in $s$-process elements through dredge-up phases during thermal 
pulses. These enriched envelopes are ejected into the ISM through stellar
winds.  Lower mass stars in the globular cluster can accrete this gas
(Thoul et al., in preparation) :
the $s$-process element enrichment occurs only in the external layers. 
With time, some of those stars can be 
ejected from the globular clusters through various 
dynamical processes, such as evaporation or disruption, and populate the 
galactic halo.  These stars account for our PopIIb.  Theoretically foreseen
for a long time (see for example Applegate, 1985; Johnstone, 1992;
Meylan and  Heggie, 1997), 
these dynamical processes dislodging stars from 
globular clusters begin to rely upon observations.
De Marchi et al.~(1999) have observed the 
globular cluster NGC~6712 with the ESO-VLT and derived its mass function.
Contrary to other globular clusters, NGC~6712 mass function shows a noticeable 
deficit in stars with masses below 0.75$M_{\odot}$.
Since this object, in its galactic orbit, has recently penetrated very deeply 
into the Galactic bulge, it has suffered tremendous gravitational shocks.  
This is an evidence that tidal forces can
strip a cluster of a substantial portion of its lower mass stars, 
easier to eject than the heavier ones.  \\
Our scenario requires that a significant fraction of the field stars now 
observed in the halo have been evaporated from globular clusters at an 
earlier epoch.
Indeed, Johnston et al.~(1999) claim
that GC destruction processes are rather efficient : a significant
fraction of the GC system could be destroyed within the next Hubble time.
However, McLaughlin (1999) and Harris and Pudritz (1994) argue that 
these various destructive mechanisms are important only for {\it low}
mass clusters.  Therefore, these clusters cannot have contributed 
much to the total field star population because of their small 
size.  As one can see, the origin of the field halo stars 
is still a much debated question. \\    

In relation with the different steps proposed above to explain the 
observations, the scenario was labeled EASE which stands for 
\it Evaporation/Accretion/Self-Enrichment \rm (Fig.~2).

\section{Model description}

\subsection{The proto-globular cluster clouds}

According to Fall and Rees (1985), PGCCs are cold ($T_c \simeq 10^4$K) 
and dense primordial gas clouds in pressure equilibrium with a hot 
($T_h \simeq 2\times 10^6$K) and diffused protogalactic medium.
As already mentioned in section 1, the PGCCs can be maintained at
a temperature of $10^4$K only if some external heat sources were present at
the protogalactic epoch.  Fall and Rees (1988) and Murray and Lin (1993) 
have proposed that the UV flux
resulting from the hot protogalactic background could be sufficient to offset
the cooling below $10^4$K in a gas with a metallicity less than 
[Fe/H]$ \simeq -$2.
As the PGCC is assumed to be made up of primordial gas
(we deal with a {\it self-}enrichment model and not a 
{\it pre}-enrichment one), we will suppose it is indeed the case.
Within the context of this preliminary model, 
the following assumptions are made : 
\begin{itemize}
\item the PGCCs are thermally supported (no turbulence or magnetic field) 
and the PGCC primordial gas obeys the perfect gas law; 
\item a PGCC of mass $M$ and radius $R$ is an isothermal sphere in hydrostatic
equilibrium, hence its density profile $\rho (r)$ scales as $r^{-2}$ : 
\begin{equation}
\rho(r) =\frac{1}{4\pi}\frac{M}{R} r^{-2}.    
\end{equation}
\end{itemize}
The requirement of pressure equilibrium at the interface between the 
cold and hot phases leads to 
\begin{equation}
P(R)=\frac{k T}{\mu m_h} \times \rho(R)=P_h   
\end{equation}
where $\mu$, $T$ and $P_h$ are respectively the mean molecular weight 
($\mu$=1.2), the temperature of the PGCC and the pressure of the 
hot protogalactic medium confining the PGCC.  Equations (1) and (2) 
lead to a relation between the radius and the cubic root of the 
mass of the PGCC :
\begin{equation}
R_{100}=\chi M_6^{1/3};
\end{equation}
subscripts ``100'' and ``6'' mean that 
radius and mass are respectively in units of 100 parsecs and
$10^6 M_{\odot}$.
Upon the assumption of a temperature of $10^4$K for the cold clouds, 
$\chi$ is defined by 
\begin{equation}
\chi  ={\left(\frac{3.7 \times 10^{-12}}{P_h}\right)}^{1/3}  
\end{equation}
where $P_h$ is expressed in dyne.cm$^{-2}$. 
The mass of the PGCC is the mass of a pressure-truncated isothermal 
sphere of gas in hydrostatic
equilibrium and is given by (McLaughlin and Pudritz 1996b) : 
\begin{equation}
M=\sqrt{\frac{2}{\pi}}\left(\frac{kT}{\mu m_h}\right)^2G^{-3/2}{P_h}^{-1/2}.
\end{equation} 
For a temperature of $10^4$K, Eq.~(5) becomes
\begin{equation}
M_6=1.1 \times 10^{-5}{P_h}^{-1/2}.
\end{equation}
From equations (1), (3) and (6), the mass, the radius and the density 
profile of a PGCC bounded by a given pressure is completely determined.
From this, we are able to find the expansion law for  a supershell 
sweeping a PGCC.
Contrary to Dopita and Smith (1986) who used the 
\it current \rm gravitational potential of the Galaxy to evaluate $\chi$,
our determination is based on protogalactic conditions.

\subsection{The formation of the first generation}

It is well known that star formation can only occur in the coolest and 
densest regions of the ISM. 
We assume that the UV external heating provided by the hot 
protogalactic background is shielded by the bulk of the PGCC gas.
Therefore, $\rm H_2$ formation and thermal cooling are assumed
to occur only near the center 
of the PGCC :  the formation of the star first generation takes 
place in the PGCC central area, the densest and the coolest regions 
of the cloud. 
For  the value of the SFE, i.e. the
ratio between the mass of gas converted into stars and the total mass of gas,
we refer to Lin and Murray (1992).  Their computation shows that the early star
formation in protogalaxies  was highly inefficient, leading to a SFE not higher
than one percent. The mass spectrum is described with the following
parameters~: 
\begin{itemize} 
\item[1. ] the slope of the initial mass function (IMF) $\alpha$, 
\item[2. ] the lower $m_{l1}$ and upper $m_{u}$ mass limits of the spectrum, 
\item[3. ] the mass of the least massive supernova progenitor $m_{l2}$, 
\item[4. ] the mass of the least massive supernova progenitor contributing
to the PGCC self-enrichment $m_{l3}$ ($>m_{l2}$).
\end{itemize}
The distinction between points 3 and 4 lies in the fact that the least massive
supernova progenitors (9 $<$ M $<$ 12$M_{\odot}$) have very thin shells of 
heavy elements when they explode 
and contribute only very slightly to the chemical enrichment even if they are 
numerous.  However, their dynamical impact on PGCC must be taken into account
(stars with masses between $m_{l2}$ and $m_{u}$ end their live as supernovae,
but only the ones with masses between $m_{l3}$ and $m_{u}$ participate 
to the chemical enrichment).

\subsection{Supershell propagation}

The model of Castor et al.~(1975) primarily describes the evolution of a 
circumstellar shell driven by the wind of an early-type star.
Their study can be extented to multiple supernova shells if the supernova 
progenitors (the first generation massive stars) are closely associated.
In this case, all supernova shells will merge into a single supershell 
propagating 
from the center to the edge of the PGCC.  Following the remarks of the previous
section, we assume that this is indeed the case. \\
The blast waves associated with the first supernova explosions sweep the PGCC 
material in a thin, cold and dense shell of radius $R_s$ and velocity 
$\dot{R_s}$. This shell surrounds a hot and low-density region, the bubble, 
whose
pressure acts as a piston driving the shell expansion through the unperturbed
ISM.  The following equations settle the expansion law $R_s(t)$ of the shell
during its propagation in a given PGCC
(Castor et al., 1975; Brown et al., 1995) :

\begin{itemize}

\item[1. ] The supernova explosions add energy to the bubble at a rate
$\dot{E_o}$ and the dominant energy loss of the bubble comes from work
against the dense shell, hence the total energy of the bubble $E_b$ 
is the solution from    
\begin{equation}
\dot{E_b}=\dot{E_o}-4\pi {R_s}^2 P_b \dot{R_s}.
\end{equation}
\item[2. ] The internal energy $E_b$ and the pressure $P_b$ of the bubble are
 related through 
\begin{equation}
\frac{4\pi}{3} {R_s}^3 P_b = \frac{2}{3} E_b.  
\end{equation}
\item[3. ] The motion of the supershell obeys Newton's second law 
\begin{equation}
\frac{d}{dt} [M_s(t) \dot{R_s}(t)] = 4\pi {R_s}^2 (P_b-P_{ext})
-\frac{{GM_s}^2(t)}{2{R_s}^2(t)}  
\end{equation}
where $M_s(t)$ is the mass of the shell at time $t$ and  
$P_{ext}$ is the pressure in the unperturbed medium just outside the shell
($P_{ext} = P(R_s(t))$,
which depends on the temperature of the isothermal cloud.
\item[4. ] Knowing the density profile of the PGCC, we obtain the mass 
swept by the shell at time $t$ 
\begin{equation}
M_s(t) = \frac{M}{R} R_s(t)
\end{equation}
where $M$ and $R$ are the mass and radius of the PGCC. 
\end{itemize}
If, as assumed by McCray (1987), the energy injection rate 
$\dot{E_o}$ supplied by the supernova explosions is constant in time,
then an analytical solution of the type $R_s(t) = v \times  t$ describes
the shell motion.  In this expression, $v$ is the constant speed of the shell
during its progression through the PGCC.  For an isothermal cloud of 
temperature $T$
and mean molecular weight $\mu$, relations (7-10) show that $v$ depends on the 
PGCC mass and radius and on the supernova rate :  
\begin{equation}
\left[3v^3+3\left(\frac{kT}{\mu m_H}+\frac{GM}{2R}\right)v\right]
=2\dot{E_o}\frac{R}{M}.
\end{equation}
Since, the mass and the radius of the cloud are determined by 
Eq.~(3),(4) and (6), the parameters of the problem 
are the pressure confining the PGCC and the supernova rate. 
Therefore, Eq.~(11) successively becomes 
\begin{equation}
[v_{10}^3+(0.7+0.2\chi ^{-1} M_6^{2/3} )v_{10}]=3.3 
\frac{NE_{51}}{\Delta t_6} \chi M_6^{-2/3},
\end{equation}
\begin{equation}
v_{10}^3+1.4v_{10}=\frac{NE_{51}}{\Delta t_6}.
\end{equation}
In Eq.~(12) and (13), the bulk of the gas is assumed to be at a temperature
of $10^4$K, the mean molecular weight $\mu$ is set to 1.2,
the expansion speed $v_{10}$ is in units of 10km.$\rm s^{-1}$ (the sound speed 
value in a medium of this temperature), $M_6$ is the mass of the cloud in 
units of millions solar masses, N is the total number of SNeII, 
$E_{51}$ is the energy provided by each Type II supernova explosion  
in units of $10^{51}$ ergs and $\Delta t_6$ is the 
supernova phase duration expressed in millions years.  Equation (13)
shows that the velocity of the shell depends on the number of SNeII 
and is independent on the hot protogalactic background pressure.

\section{Level of self-enrichment}

We have assumed that stars form in the central regions of
a given PGCC. This first generation of stars is spontaneous, i.e. not 
triggered by any external event, a shock wave for instance.  This results in a 
low SFE.  After a few millions years, 
the massive stars ($9M_{\odot}<M<60M_{\odot}$) end their lives as SNeII.  
The consequences are :
\begin{itemize}

\item[1. ] An enrichment of the primordial cloud.  This is the 
\it self-enrichment \rm phenomenon : the cloud has produced its own source 
of enrichment.  As we will show below, this self-enrichment can 
explain the globular cluster metallicities.

\item[2. ] A trigger of blast waves sweeping all the ISM in an expanding 
shell.  The high density in the shell favours the birth of a second generation
of stars with a higher SFE than for the first one.
 
\end{itemize}

\subsection{Dynamical constraint}

As the energy released by one SNII is typically $10^{51}$ ergs
($E_{51}$=1), the energy of a few supernova explosions and the
gravitational binding energy of a PGCC are of the same order of magnitude.
This is the major argument used against self-enrichment.
Actually, it is often argued that successive supernovae will disrupt the 
proto-globular cluster cloud.  
To test whether this idea is right or not, we can compare the gravitational 
energy of the cloud to the kinetic energy of the expanding shell, supplied 
by the supernova explosions, when it reaches the edge of the cloud.
Indeed, taking the following criterion for disruption :  
\begin{equation}
\frac{1}{2} v^2 = \frac{GM}{R}
\end{equation}
where $v=\dot{R_s}(t_e)$ and $t_e$ is the time at which the shell reaches 
the edge of the cloud, i.e. the time at which the whole cloud has been 
swept-up in the shell, we have 

\begin{equation}
t_e=\frac{R}{v}.
\end{equation}
From equations (3), (12) and (14),  we get

\begin{eqnarray}
0.8 \chi ^{-3/2} M_6 + 0.6 \chi ^{-1/2} M_6^{1/3} + 
0.2 \chi ^{-3/2} M_6  = \nonumber \\  
3.3 \chi M_6^{-2/3} \frac{N E_{51}}{\Delta t_6}. 
\end{eqnarray}
In this equation, N is the maximum number of SNeII a PGCC of 
mass M can sustain without being disrupted.
The second and third terms in the left-hand side account for the 
decelerating effects produced by the external pressure and the shell gravity 
respectively.
Again, using Eq.~(4) and (6), we obtain
\begin{equation}
NE_{51}=201
\end{equation}
where the duration of the SNII phase is set to 30 millions years
($\Delta t_6 = 30$).  Therefore, a PGCC, here described by a 
pressure-truncated isothermal (T $\simeq 10 ^4$K) sphere of gas in
hydrostatic equilibrium, can sustain {\it numerous} SNII explosions.

\subsection{Chemical constraint}

The first generation of stars is not triggered by any event (shock wave 
for instance) and therefore the SFE is likely to be be very low.  
The typical halo metallicity of [Fe/H]$\sim -$1.6 must however be 
reached despite this low first 
generation SFE.  We now show that this is indeed the case. 
We compute the relation between the mass of primordial gas 
and the number of first generation supernovae for two different 
cases.  In the first case, we assume a given SFE (plain curve in Fig.~3), 
while in the second one, we impose the final metallicity (dashed curves in 
Fig.~3).  In what follows, all the masses are 
in units of one solar mass. \\

The mass distribution of the first generation of stars obeys the 
following power-law IMF,
\begin{equation}
dN = k m^{- \alpha} dm 
\end{equation}
where $dN$ is the number of stars with masses between $m$ and $m+dm$, 
$\alpha$ is the
power-law index and $k$ is a coefficient depending on the total mass of the 
first generation. \\

The PGCC (of total mass M) is made up of two components, one being
the stars (total mass $M^{*}$), the other being the remaining gas, i.e.
the gas not consumed to form the first generation stars (total mass 
$M^{gas}$).  We can therefore write 
\begin{equation}
M=M^{*}+M^{gas}. 
\end{equation}

The mass $m_z$ of heavy elements ejected in the ISM by a supernova
whose progenitor mass is $m$ is approximatelly given by (Woosley and Weaver, 
1995)
\begin{equation}
m_z = 0.3m-3.5, 
\end{equation}
where 12$<m<$60. \\

The total mass $M^{*}$ of first generation stars, the number $N$ of 
SNeII and the total mass $M_z$ of heavy elements ejected by all 
supernovae
are given by 
\begin{equation}
M^{*}=\int_{m_{l1}}^{m_u} {k m^{-\alpha + 1}}\,dm, 
\end{equation}

\begin{equation}
N=\int_{m_{l2}}^{m_u} {k m^{-\alpha}}\,dm, 
\end{equation}

\begin{equation}
M_z = \int_{m_{l3}}^{m_u} {k m^{-\alpha} m_z}\,dm 
\end{equation}
where $m_u$ is the upper mass limit, $m_{l1}$ is the lower mass limit for the 
IMF, $m_{l2}$ is the lowest star mass leading to a  
SNII and $m_{l3}$ is the lowest star mass producing heavy elements. \\

If the mass $M_z$ of heavy elements is mixed into a mass $M^{gas}$ of 
primordial gas, the final metallicity [Fe/H] of the gas cloud is defined by 

\begin{equation}
M_z = Z_{\odot} 10^{[ \rm Fe/H]} M^{gas} 
\end{equation}
\noindent
where $Z_{\odot}$ is the solar heavy element mass abundance.\\

\begin{figure}
\vspace*{-1.5cm}
\begin{center}
\leavevmode 
\epsfxsize= 8.5 cm 
\epsffile{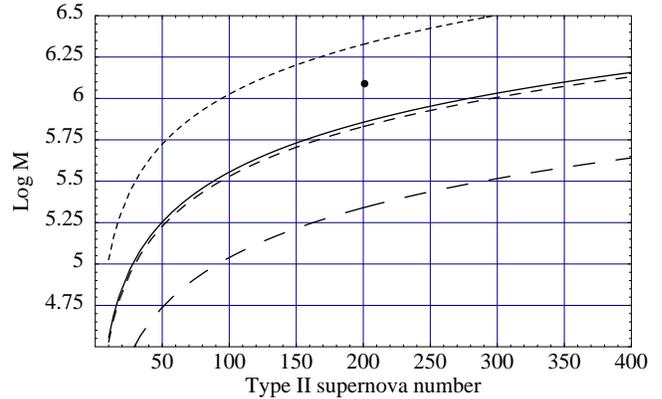}
\vspace*{-1.5cm}
\caption{Relations between the mass of the PGCC and the number of  
SNeII exploding in its central regions for a given SFE = 0.01
(plain curve), three self-enrichment levels (dashed curves), from up to 
bottom, 
[Fe/H]$=-2, -1.5, -1$.  The point represents the dynamical constraint 
described in the text for a hot protogalactic background pressure of 
8 $\times$ 10$^{-11}$dyne.cm$^{-2}$}
\end{center}
\end{figure}
\noindent
Given either the SFE ($=M^*/M$) or the final metallicity [Fe/H], we can solve 
equations (18) to (24) to obtain the relationships between N and M 
in both cases.
In Fig.~3, we have used  
$\alpha =$ 2.35 (Salpeter 1955), 
$m_{l1}=3(M_{\odot}$),
$m_{l2}=9(M_{\odot}$),
$m_{l3}=12(M_{\odot}$),
$m_{u}=60(M_{\odot}$), 
$M^*/M$ = 0.01 and three different values for the final metallicity :
[Fe/H]$=-2, -1.5$ and  $-$1. 
We will justify these values for the IMF parameters in the next section.  
The point represents the dynamical constraint (17), namely
the maximum number of SNeII a PGCC can sustain without disruption.
The PGCC mass is set by the value of $P_h$ (Eq.~6).
The points located left/right  
correspond respectively to stable/unstable PGCCs.  The last ones 
can't give rise to GCs as, following Eq.~(14), they are disrupted.
Similarly, the clouds with masses and number of SNeII above/under the 
plain curve (SFE=0.01) correspond to clouds with SFEs smaller/greater 
than one percent.  The three dashed curves represent relations between 
the primordial gas mass and the number of SNeII at a given metallicity.  
From top to bottom, these metallicities are [Fe/H]=$-$1, $-$1.5, and $-$2.
We deduce from Fig.~3 that we can achieve a metallicity 
[Fe/H]$\sim -$1.5 with an SFE of one percent. 
Hence, \it  the mean metallicity of the 
galactic halo can be explained by the activity of a spontaneous first 
generation of stars, even with a low SFE\rm.  
Furthermore, the graphical 
comparison of the dynamical constraint with the set of curves corresponding to
the three different metallicities shows that \it the dynamical constraint is 
also compatible with the metallicities of some halo globular clusters \rm.
For instance, under a hot background pressure of 8$\times$10$^{-11}$
dyne.cm$^{-2}$, the mass of a PGCC is 1.2 $\times$ 10$^6$ M$_{\odot}$
and the metallicity that can be achieved through self-enrichment is $-$1.75.
Table 1 illustrates the sensitivity of the model to the choice of $P_h$.   
\begin{table}[h]
\caption[]{Dependence of [Fe/H] on $P_h$}
\begin{center}
\begin{tabular}{|c|c|c|c|} \hline 
$P_h [dyne.cm^{-2}]$ & $log_{10} M_{c}/M_{\odot}$ & [Fe/H] 
\\ \hline
$10^{-9}$ &  5.5 & -1.2 \\
$10^{-10}$ &   6.0 & -1.7 \\
$10^{-11}$ &   6.5 & -2.2 \\ \hline
\end{tabular}
\end{center}
\end{table}

Our main conclusion is therefore that galactic halo 
globular cluster metallicities might be consistent with the process of 
self-enrichment, both from dynamical and chemical points of view. \\
It is clear that the metallicities presented in Tab.~1 
are not compatible with the disk 
globular cluster metallicities.  However,
the metallicity distribution of the Milky Way GCs exhibits 
two distinct peaks, suggesting the existence of two subpopulations of GCs,
a metal-poor one and a metal-rich one.  
The disk population might be a second generation of GCs 
formed out of gas where significant {\it pre}-enrichment had occured,
due to the early formation of the galactic halo.  
For these clusters, there is no characteristic mass in the Fall and Rees 
sense : the temperature can decreased significantly below $10^4$K as 
cooling is provided by the heavy elements.  Therefore,
others mecanisms must be invoked.  For 
disk GC formation, see Burkert {\it et al.}~(1992), 
Ashman and Zepf (1992).

\section{Discussion}
  
\subsection{The IMF of the first generation} 

The first generation of stars forms out of a gas very poor, or even free,
in heavy elements.  We now examine the consequences for our model.

\subsubsection{The shape of the IMF in a metal deficient medium}

As already mentioned in section 3.2, we assume that the gas 
temperature can decrease significantly below 10$^4$K in the 
central regions of the PGCC only.  
We now focus on what might happen in this central region where 
star formation is expected.  \\
According to Larson (1998), the Jeans scale can be identified 
with an intrinsic scale in the star formation process.   
It is defined as the minimum mass that a gas clump must have in order for 
gravity to dominate over thermal pressure (although the {\it thermal}
Jeans mass is not universally accepted as relevant to the present-day
formation of stars in turbulent and magnetized molecular clouds).  
It scales as
\begin{equation}
M_J \propto T^2 P^{-1/2}
\end{equation}
where T is the temperature of the clump and P is the ambiant pressure 
in the star forming region (here the PGCC central area) surrounding 
the clump. 
There is therefore a strong dependence on the 
temperature.   At $T$ smaller than $10^4$K,  
heavy elements and dust grains provide the most efficient cooling mechanism 
in present-day clouds while in primordial clouds the cooling process is 
likely to be governed by 
molecular hydrogen, the only available coolant in a zero metallicity medium.
However, H$_2$ cannot cool the gas much below 100K while heavy elements can 
make the temperature as low as 10K. 
Therefore, we expect to have higher clump temperatures and therefore higher 
star formation mass scales in primordial gas than in current
star forming regions.
The IMF has in all likelihood evolved with metallicity and 
therefore with time, probably favouring higher masses in the early Universe.
To take into account this time variability, Larson (1998) has proposed
two alternative definitions for the IMF, where the slope is fixed 
(rather than variable 
as usually assumed in earlier analyses) and where the variable parameter 
is the mass scale $m_1$ :  

\begin{equation}
\frac{dN}{d\log m} \propto m^{-1.35} exp(-m_1/m)
\end{equation} 

and

\begin{equation}
\frac{dN}{d\log m} \propto \left(1+ \frac{m}{m_1}\right)^{-1.35}.  \\
\end{equation} 
These IMFs have a universal 
Salpeter form at large masses but depart from this power law below a 
characteristic mass that can vary (Fig.~4).
Recent observational results suggest that the IMF has indeed a 
universal Salpeter slope at the upper end and that any variability should 
be confined to the lower end (Haywood 1994, Meyer et al.~1999). 
Contrary to the Salpeter IMF which raises monotically 
with decreasing stellar mass, the current observed IMF in the solar 
neighbourhood may be roughly flat at the lower end or even may peak at a mass 
around $0.25M_{\odot}$ and decline into the brown dwarf regime.
The uncertainty about the behaviour at the lower mass end comes from the fact 
that the IMF below $0.1M_{\odot}$ remains very poorly known.  
Whatever the exact solution, a single power-law Salpeter IMF can be ruled 
out at masses lower than $1M_{\odot}$, the IMF slope
becoming clearly much smaller than the Salpeter value.
The Larson IMFs (21) and (22) are in agreement with the observations since 
they reproduce the Salpeter's behaviour for large
masses while they allow a large range of possibilities  
at the lower mass end depending on the value adopted for the mass scale $m_1$.
The essential result is to alter the relative 
number of low-mass stars compared with the number of more massive ones. 
If the mass scale was higher at early times, relatively fewer low-mass stars 
would have been formed at these times. \\

\begin{figure}
\vspace*{-1.5cm}
\begin{center}
\leavevmode 
\epsfxsize= 8.5 cm 
\epsffile{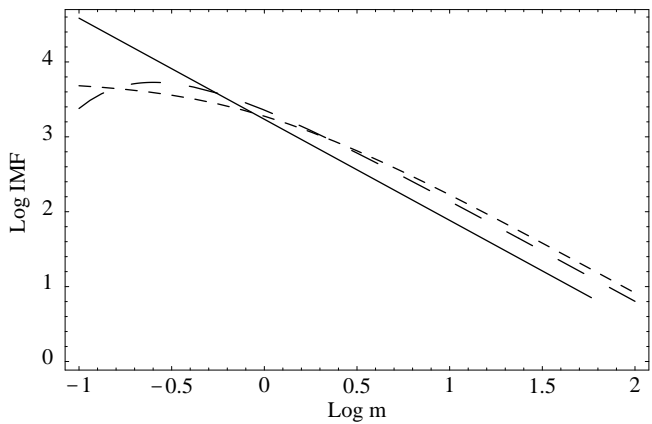}
\vspace*{-1.5cm}
\caption{Comparison of Salpeter IMF (plain curve) versus Larson modified IMFs
(dashed curves).
The results have been normalized in such a way that the mass integral is the 
same for the three IMFs}
\end{center}
\end{figure}

Furthermore, Nakamura and Umemura (1999) have studied the cooling properties
of molecular hydrogen. From this, they have estimated that, in a 
zero metallicity medium, the lowest mass star allowed to form is a 3$M_{\odot}$
star.  In other words, instead of any flattening or decline of the 
IMF, there is in this case a sharp cut off at the low mass end.

\subsubsection{The star formation efficiency}
The star formation efficiency is an important parameter, since   
the final mass fraction of heavy elements released
in the primordial medium depends linearly on it.
How stars are formed out of a gaseous cloud is still poorly known
and it is not easy to estimate the value of the SFE.
One of the most crucial step in the star formation process is 
the creation of molecular hydrogen (Lin and Murray 1992).  H$_2$
cooling results in a rapid burst of star formation which continues until 
massive stars have formed in sufficient number to reheat the surrounding gas.
The massive stars produce a UV background flux which  
destroys the molecular hydrogen by photodissociation and shuts down further 
star formation.  
Applying this principle of \it self-regulated star formation by negative 
feed back \rm to a protogalactic cloud, Lin and Murray (1992) have calculated 
the UV flux necessary to destroy the molecular hydrogen and the 
required number of massive stars to produce this UV flux.  
Finally, the mass and the  
SFE of this first generation of stars in the protogalaxy are estimated.
They find a value of the order of one percent. 
Under the asumed IMF, an SFE of one per cent corresponds to a metallicity
of $-$1.5 (see Fig.~3).

\subsubsection{The upper limit of the mass spectrum}
\begin{figure}
\vspace*{-1.5cm}
\begin{center}
\leavevmode 
\epsfxsize= 8.5 cm 
\epsffile{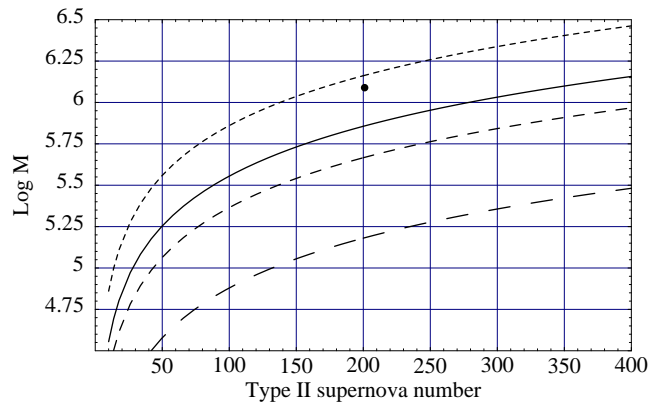}
\vspace*{-1.5cm}
\caption{Sensitivity to the mass spectrum upper limit.  The curves shown here 
are similar to those in Fig.~3.  Compared to these previous results, 
lower metallicity is reached if the upper mass 
limit of stars contributing to the enrichment of the ISM is set to 
40M$_{\odot}$ instead of 60M$_{\odot}$}
\end{center}
\end{figure} 
In the model we have adopted in section 4, the mass of heavy elements released
by a star is proportional to its total mass (see Eq.~(20)).  
However, above a critical mass $m_{BH}$, a star can form a black hole without 
ejecting the heavy elements it has processed.  This critical mass is given by 
$m_{BH}=50 \pm$10$M_{\odot}$ (Tsujimoto et al., 1995).  
Moreover, Woosley and 
Weaver (1995) have shown that zero initial metallicity stars have a final 
structure markedly different from solar metallicity stars of the 
same mass.  The former ones are more compact and larger amounts of 
matter fall back after ejection of the envelope in the SN explosion.  
In this case, more heavy elements are locked in the remnant left by the most 
massive stars.  In order to evaluate the consequences for our model, we 
now define two different upper mass limits for the IMF.  The mass of the most 
massive supernova contributing to the  
enrichment of the ISM is chosen to be $m_{u1}=40M_{\odot}$.  
But the more 
massive stars will have a dynamical impact on the ISM and 
contribute to the trigger and the early expansion of the supershell,
even though they will not contribute to the self-enrichment. 
All stars with masses
between   $m_{l2}$ and $m_{u2}$ end their lives as supernovae,
but only the ones with masses between $m_{l3}$ and $m_{u1}$ contribute 
to the PGCC self-enrichment.
We adopt $m_{u2}=60M_{\odot}$ as the mass of the most massive supernova 
progenitor.  This value is the same as in section 4.2 and therefore 
the plain curve in Fig.~3 (given SFE) is not modified.    
Keeping the same SFE, if we decrease $m_{u1}$ from 60$M_{\odot}$ 
to 40$M_{\odot}$, 
there is a reduction of $\sim$ 0.2 dex in the final metallicity (Fig.~5).
An IMF with a Salpeter slope doesn't favour at all the highest mass stars 
and these 
stars are quite rare compared to the less massive supernovae.  
This explains why the final metallicity is not strongly dependent on 
the value  of $m_{u1}$.

\subsubsection{The lower limit of the mass spectrum}
We have adopted the point of view of Nakamura and Umemura (1999) who
assume that there is a sharp cutoff at the lower mass end of the IMF 
($m_{l1}=3M_{\odot}$).  If we consider a pure Salpeter IMF, this 
parameter is very critical.  Indeed, if we take $m_{l1}=0.1M_{\odot}$ 
instead of 3, while keeping the SFE unchanged, the mass of heavy elements 
ejected by SNeII is decreased by a factor of four, leading to 
a decrease in metallicity of 0.6 dex.  
The Larson's modified IMFs provide less sensitive results.

\subsubsection{The slope of the IMF}
Following Larson (1998), we have used the Salpeter value.
Changing the slope of the IMF will have the same consequences as changing 
the value of the lower mass cut off.  
If we decrease the slope, we will increase the ratio of 
high mass stars over low mass stars.
The same result can be obtained by increasing the lower mass cut off of 
the mass spectrum.  

\subsection{Observational constraint : the RGB narrowness}

With the exception of $\omega$ Cen, and perhaps M22, galactic globular 
clusters share the 
common property of a narrow red giant branch, indicative of chemical 
homogeneity within all stars of a given globular  cluster.  
This observational property is also  
often used as an argument against self-enrichment.  
We now show that self-enrichment and a narrow RGB are in fact compatible. 

\subsubsection{The mass of the first generation stars}

\begin{figure}
\vspace*{-1.5cm}
\begin{center}
\leavevmode 
\epsfxsize= 8.5 cm 
\epsffile{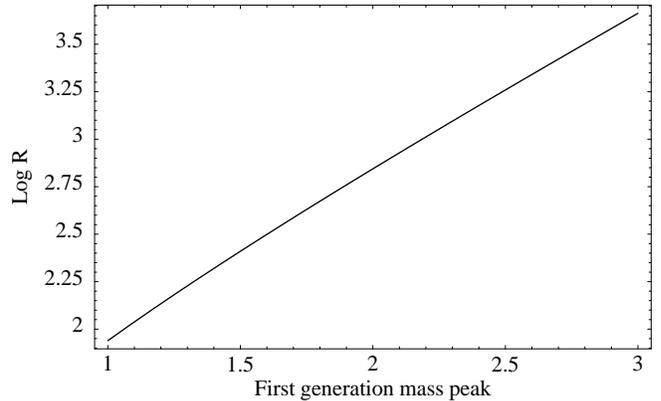}
\vspace*{-1.5cm}
\caption{Ratio R of the numbers of low mass stars in the two generations 
assuming that the second SFE is ten times higher than the 
first one.  The second generation mass peak is $0.25M_{\odot}$ while the 
first generation mass peak is left as a parameter}
\end{center}
\end{figure} 

If we adopt the lower mass limit for the mass spectrum of initially metal-free
stars proposed by Nakamura and Umemura (1999), we get rid of one of the 
major arguments against self-enrichment scenarios in globular clusters : 
the existence of two distinct generations of stars with clearly different 
metallicities.  Indeed, if the first generation of stars is biased towards high
mass stars as previously suggested, these stars are no more 
observed today and the current width of the RGB is not affected. \\
However, following Larson (1998), we could also allow low mass star 
formation
during the first phase \it but under a different mass-scale than today\rm.
Using the IMF given by equation (26), we have computed the ratio between
the currently observed numbers of low-mass ($0.1M_{\odot}<M<0.8M_{\odot}$) 
stars, which are produced in both generations.  According to
Brown et al.~(1995), the second generation of stars will form a bound globular 
cluster if its SFE is at least 0.1.  We therefore assume a value of 10 for
the ratio SFE(2$^{nd}$ generation)/SFE(1$^{st}$ generation). 
The second generation mass scale $m_1$ (see Eq.~(26)) has been fixed to 
$0.34M_{\odot}$ 
to match the solar neighborhood potential peak located at $m=0.25M_{\odot}$
(see section 5.1.1).
The mass peak of the first generation stars is left as a parameter 
and we allow it to vary between 1 and 3$M_{\odot}$.
The ratio R of the number of second generation low mass stars to
the number of first generation low mass stars is shown in Fig.~6.  
For one metal-free star, 
the number of second generation stars lies between 100 and 
4000 depending on the first generation mass peak. 
Thus, even if low mass stars were formed in the metal-free 
PGCC, their relative number observed today is so small that the existence of 
the first generation is not in contradiction with the RGB narrowness 
observed on globular cluster CMDs.

\subsubsection{The time of formation of the second generation}

Another controversial point about self-enrichment concerns the ability of the 
shell to mix homogeneously the heavy elements with the primordial gas.
If the mixing is not efficient,
inhomogeneities will be imprinted in the second 
generation stars formed in the shell and will show up as a broader red 
giant branch in CMDs, contrary to observations.  
Brown et al.~(1991) have established 
that the accretion time by the blast wave propagating ahead of the  
shell is one to two orders of magnitude larger than the mixing time due to 
post-shock turbulence in the shell.  In other words, the material swept by 
the shell is more quickly mixed 
than accreted and the post-shock turbulence insures supershell homogeneity.  
But even if the supershell is chemically homogeneous
\it at a given time\rm, the chemical composition varies with time : 
metallicity is increasing as more and more supernova explosions occur 
at the center of the bubble.  So, one can argue that the second generation 
will not be homogeneous : stars which are born early will be more metal-poor 
than stars which are born later, when self-enrichment has gone on.  
However, shell fragmentation into molecular clouds,
 in which second generation stars 
will form, cannot take place too early, at least not before the death of 
all first generation O stars.  Indeed, these ones 
are the most important UV flux emitters and, as such, prevent the formation of
molecular hydrogen.   \\ 
It is very interesting to plot the increase of metallicity versus time
when the shell has emerged in the hot protogalactic medium (all the cloud 
material has been swept in the shell whose mass is now a constant).
For simplicity, we have assumed that the ejecta of one supernova mix
with all the PGCC gas before the next supernova explosion.
In Fig.~7, the parameter values are the same as in section 4
(same IMF, N=201, $P_h$=8 $\times$ 10$^{-11}$ dyne.cm$^{-2}$) and the relation
between the mass of a SNII progenitor and its lifetime on the Main Sequence
is given by $\tau_{MS}\simeq 3\times 10^7 yr[M_*/(10M_{\odot})]^{-1.6}$
(Mc Cray, 1987).   
In this case, the shell reaches the edge of the cloud 2.3 millions years 
(explosion of the 35$M_{\odot}$ supernovae) 
after the first explosion.
We see in Fig.~7 that after a rapid increase in metallicity, as expected for 
a metal-free medium, the increase in metallicity slows down and saturates.  
After 9 millions years, when all stars more massive than 19$M_{\odot}$ 
have exploded, the further metallicity increase is less than 0.1 dex, 
the upper limit of the RGB metallicity spread.  
Therefore, there is no conflict between a 
self-enrichment scenario and the RGB narrowness \it if the second generation 
of stars is  born after this time \rm .  Even if supernova 
explosions still occur, the self-enrichment phase has ended.  This point was 
already underlined by Brown et al.~(1991) from a dynamical point of view.
\begin{figure}
\vspace*{-1.5cm}
\begin{center}
\leavevmode 
\epsfxsize= 8.5 cm 
\epsffile{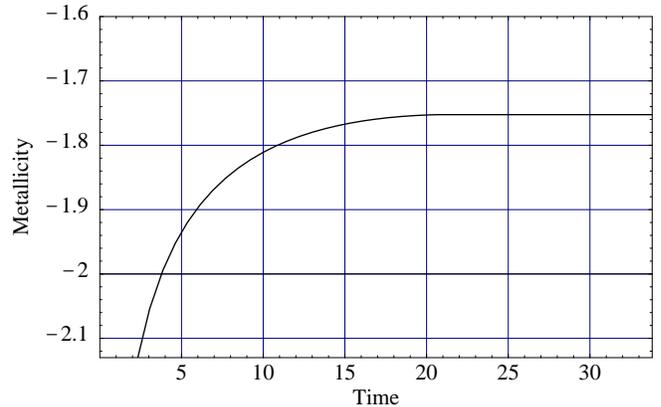}
\vspace*{-1.5cm}
\caption{Progressive enrichment due to Type II supernovae.  Time is expressed 
in millions years}
\end{center}
\end{figure}

\section{Conclusions}

We have investigated the possibility that globular clusters have undergone 
self-enrichment during their evolution.
In our scenario, massive stars contribute actively
to the chemical enrichment and to the gas dynamics in the early Universe.
When a stellar system is formed, supernovae enrich the remaining gas
in such a way that the next generation of stars is more metal-rich than the
first one.  In this paper, we assume the birth of a first generation of 
stars in the 
central areas of PGCCs.  When the massive stars end their lives, the 
corresponding SNeII explosions trigger the expansion of a 
spherical shell of gas, where the PGCC primordial gas and the heavy elements 
ejected by supernovae get mixed.  Because of the dynamical impact of supernova
shock waves on the ISM, the gas is compressed into a dense shell and this high 
density favours the birth of a second generation of stars with a higher SFE.
This scenario of triggered star formation is now confirmed by observational 
examples in the disk of our Galaxy and in irregular galaxies. 
The second generation stars formed in these compressed layers of gas are the 
ones 
we observed today in GCs.  Others authors have proposed scenarios where 
these stars are also formed in triggered events, namely in gas layers 
compressed by shock waves, but the origin of the trigger is different.  
For instance, following Vietri and Pesce (1995) and Dinge (1997), the 
propagation of shock waves in the cloud could be respectively promoted by 
thermal instabilities inside the cloud or cloud-cloud collisions.  
Thus, in these 
scenarios, there is no first generation massive stars as shock wave sources :
this is the major difference between our scenario and theirs. \\

It has long been thought that PGCCs were 
not able to sustain SNeII explosions because of the associated
important energetic effects on the surrounding ISM.  
In this paper, we have shown that this idea may not be true.
For this purpose, the criterion for disruption proposed by Dopita and Smith 
(1986) was used.  Nevertheless, we have extended it to more general 
conditions.  Owing to the shell motion description proposed by Castor et 
al.~(1975), the spacing in time of the supernovae explosions was taken into 
account.  Also, we have not considered a tidal-truncated cloud as Dopita and 
Smith did but a pressure-confined one, which is certainly more suitable to 
protogalactic conditions.  With this model, we have computed the speed of 
propagation of the shell through the PGCC for a given supernova rate
and a given external pressure.
We have demonstrated that a PGCC can sustain many supernova explosions.  
Moreover, the dynamical upper limit on the number of SNeII
is compatible with an enrichment of the primordial gas clouds to 
typical halo globular cluster metallicities. 
This conclusion is quite robust to changes in 
IMF parameters.  Our result depends on the hot protogalactic pressure 
confining the PGCC and implies therefore a relationship between the 
metallicity and the radial location in the protogalaxy.  We have also pointed 
out that a scenario which involves \it 
two distinct \rm generations of stars is not in contradiction with the RGB 
narrowness noticed in CMDs of nearly all galactic globular clusters providing 
that the birth of the second generation of stars is not triggered before 
the 19$M_{\odot}$ supernova explosions have occurred.  
In a forthcoming paper, the correlations expected from this 
self-enrichment model will be deduced and compared to the observational data 
of the galactic halo GCs.  

\section*{Acknowledgements} 
We are very grateful to Dean McLaughlin, whose suggestions as referee of 
this paper have resulted in several improvements over its original version. 
This work has been supported by contracts ARC 94/99-178 
``Action de Recherche Concert\'ee de la Communaut\'e Fran\c{c}aise de 
Belgique" and P\^ole d'Attraction Interuniversitaire
P4/05 (SSTC, Belgium).

\end{document}